\documentclass[reprint, superscriptaddress,preprintnumbers,nofootinbib,aps]{revtex4-1}
\pdfoutput=1

\usepackage{amsmath,amssymb,physics,tensor}

\usepackage{graphicx}
\graphicspath{{figures/}} 
\usepackage[caption=false]{subfig}

\usepackage{tabularx}

\normalsize\usepackage{booktabs} 

\usepackage{xcolor}
\definecolor{navyblue}{rgb}{0.0, 0.0, 0.5}
\definecolor{tealgreen}{rgb}{0.0, 0.6, 0.6}
\definecolor{seablue}{rgb}{0.2, 0.5, 0.8}

\usepackage[colorlinks=True]{hyperref}
\hypersetup{linkcolor = navyblue, citecolor= tealgreen,urlcolor = navyblue}

\newcommand{\ba}{\begin{eqnarray}}
\newcommand{\ea}{\end{eqnarray}}
\newcommand{\be}{\begin{equation}}
\newcommand{\ee}{\end{equation}}
\def\nn{\nonumber}

\usepackage[normalem]{ulem}

\begin{document}

\author{Johanna N.~Borissova}\email{jborissova@perimeterinstitute.ca}
\affiliation{Perimeter Institute for Theoretical Physics, 31 Caroline Street North, Waterloo, ON, N2L 2Y5, Canada}
\affiliation{Department of Physics and Astronomy, University of Waterloo, 200 University Avenue West, Waterloo, ON, N2L 3G1, Canada}

\title{Suppression of spacetime singularities in quantum gravity}

\begin{abstract}
	
We investigate the requirement of suppressing spacetime geometries with a curvature singularity via destructive interference in the Lorentzian gravitational path integral as a constraint on the microscopic action for gravity. Based on simple examples of static spherically symmetric spacetimes, we demonstrate that complete singularity suppression in the path integral stipulates that the action for gravity be of infinite order in the curvature.

\end{abstract}

\maketitle

\section{Introduction}\label{Sec:Introduction}

Einstein's general relativity paved the way for a century of harmony between theoretical predictions on the one hand and experimental tests of exceedingly high precision on the other~\cite{Will:2014kxa}. Yet, it fails to provide a fundamental description of gravity at microscopic scales.

The breakdown of general relativity as a fundamental theory is indicated already at the classical level by the Penrose-Hawking singularity theorems~\cite{Hawking:1970zqf}. These establish the existence of unphysical spacetime singularities in the form of geodesic incompleteness for generic classes of solutions to Einstein's field equations. The ultimate loss of predictivity occurs in regions of large local curvature where a quantum counterpart in lieu of classical gravity is needed. Attempts to quantize general relativity by means of conventional quantum field theory have infamously settled its perturbative non-renormalizability~\cite{tHooft:1974toh,Goroff:1985th,vandeVen:1991gw}. 

In the search for a non-perturbative theory of quantum gravity, various approaches including loop quantum gravity~\cite{,Ashtekar:2021kfp,Perez:2012wv}, asymptotic safety~\cite{Eichhorn:2018yfc} and causal dynamical triangulations~\cite{Loll:2019rdj}, rely on a notion of Feynman path integral over spacetime geometries and matter fields weighted by an exponentiated action. The strict definition of such a path integral for quantum gravity is influenced by many choices. In what follows we shall adopt a general viewpoint on the path integral as an object summing over elements out of a predefined configuration space. We will not require any further relevant ingredients except the set of spacetime configurations integrated over and the bare action dictating the classical dynamics. In particular, we will not carry out any explicit evaluations of a path integral, for which we would need to $a)$ restrict considerations to a given quantum-gravity model and $b)$ apply methods which typically require a priori knowledge of the stationary points of the action, as well as make preassumptions about which configurations contribute dominantly to the path integral.

Let us on these grounds consider the path integral for a purely gravitational system, where matter fields have been integrated out and where the configuration space is defined to be the full set $\{[g_{\mu\nu}]\}$ of equivalence classes of metrics modulo diffeomorphisms. Thus the domain of integration in the formal expression for the gravitational path integral~\footnote{Planck units are implied everywhere, $\hbar = c=G=1$.},
\be\label{eq:PathIntegralFormal}
Z= \int \mathcal{D}\qty[g_{\mu\nu}]\, e^{\imath S[g_{\mu\nu}]}\,,
\ee
includes regular and singular geometries. In particular, the path integral extends over geometries with a spacetime scalar-curvature singularity defined as the divergence of a polynomial invariant of the Riemann tensor.~\footnote{We adopt the conventional definition of scalar-curvature singularity as the existence of a divergent invariant constructed polynomially from the Riemann tensor, without covariant derivatives of the metric.} A prominent example is the Schwarzschild black-hole spacetime for which the Kretschmann scalar $R_{\mu\nu\rho\sigma}R^{\mu\nu\rho\sigma}\propto r^{-6}$ diverges in the limit $r\to 0$. 

As a classical solution with an unphysical curvature singularity, the Schwarzschild geometry is unlikely to provide a viable description of spacetime in the quantum-gravitational regime. Instead, it is expected that solutions to the quantum dynamics turn out to be singularity-free and that the dominant contribution to the gravitational path integral~\eqref{eq:PathIntegralFormal} comes from such regular geometries. Given that the dynamics of gravity at high energies is still unknown, selection principles are required to constrain the variety of proposed microscopic gravitational actions. 

In the following we investigate consequences of one such possible selection principle, namely, the requirement that off-shell spacetimes with a scalar-curvature singularity are fully suppressed in the Lorentzian gravitational path integral~\eqref{eq:PathIntegralFormal}.
The suppression may originate from a modified gravitational action at high energies which guarantees their destructive interference~\cite{Lehners:2019ibe, Borissova:2020knn, Lehners:2023fud}. The appropriate modification of the action is guided by the condition of a rapidly oscillating imaginary phase factor in the path integral, accomplished through a large absolute value of the action when evaluated on spacetimes with a curvature singularity. Demanding the complete suppression of off-shell singular spacetimes in the path integral imposes a selection principle for the classical and quantum dynamics in the following sense. On the one hand, the mechanism entails to suppress singular spacetimes by altering the classical action and thus the classical dynamics of gravity at high energies. On the other hand, the path integral~\eqref{eq:PathIntegralFormal}, where microscopic matter degrees of freedom have been integrated out, defines the quantum effective action from which the quantum dynamics of gravity is derived. Therefore, the idea developed originally in~\cite{Lehners:2019ibe, Borissova:2020knn} and extended in this work can be framed more broadly as {\it singularity-suppression selection-principle for the dynamics of gravity} and represents a directed route towards constraining the significant number of proposed classical and quantum actions for gravity.
 
 Under the assumption that spacetime is continuous and described by pseudo-Riemannian geometry up to and beyond the Planck scale with no fundamental ultraviolet (UV) cutoff present, in this work we derive criteria on the action input in the Lorentzian path integral, necessary to allow for a suppression of singular spacetimes through destructive interference. To that end, we analyze the effects of generic classes of local corrections to the classical Einstein-Hilbert action on the imaginary phase factor in the Lorentzian path integral~\eqref{eq:PathIntegralFormal}, for different families of static spherically symmetric geometries.
 
As a first example we introduce a simple one-parameter family of static spherically symmetric spacetimes with and without a curvature singularity depending on the value of the free parameter. Despite the metric determinant of these spacetimes being parameter-independent and uniquely specified by the determinant of the metric on $\mathcal{S}^2$, any action integral built from finitely many contracted Riemann curvature tensors $\mathcal{R}$, in the form $\mathcal{R}^n$ for $n\in \mathbb{N}$, is incapable of achieving the complete suppression of all singular spacetimes out of this family in the path integral. The limit $n\to \infty$ ensures that all singular representatives are suppressed, whereas all regular ones make a finite equiprobable contribution. There exist, however, curvature invariants with a finite number of covariant derivatives of the metric which are sufficient to achieve the suppression of all representatives with a spacetime singularity.
These observations are tied to the specific form of polynomial Riemann invariants for the first family of spacetimes.

In the next stage we investigate more generally the possibility of singularity suppression in the path integral by means of local curvature invariants $\mathcal{R}^n$ and curvature-derivative invariants of quadratic order in the curvature given by $\mathcal{R}\Box^k \mathcal{R}$ for $k\in \mathbb{N}$, where $\Box = g_{\mu\nu} \nabla^\mu \nabla^\nu$ denotes the covariant d'Alembert operator. To that end, we consider a second example of static spherically symmetric spacetimes where the two free functions obey a power law in the radial coordinate. By analyzing subcases of the resulting two-parameter family, we illustrate that neither of the above local corrections can guarantee the full dynamical singularity suppression in the path integral for any finite $n,k\in \mathbb{N}$. Notably, in certain situations the non-trivial action of the covariant d'Alembertian reduces the efficiency of the invariants $\mathcal{R}\Box^k \mathcal{R}$ as singularity-suppression terms in the action for increasing $k\in \mathbb{N}$. Instead, the full parameter space of singular geometries described by the second family of static spherically symmetric spacetimes can be reached asymptotically only by invariants built from infinitely many curvature tensors. Altogether, these results indicate that actions of infinite order in the curvature are required to achieve the complete dynamical suppression of geometries with a spacetime scalar-curvature singularity in the gravitational Lorentzian path integral.

The paper is organized as follows. Section~\ref{Sec:DynamicalSingularitySuppression} motivates the idea of suppression of spacetimes with a scalar-curvature singularity in the Lorentzian gravitational path integral as a selection principle for the dynamics of gravity. Section~\ref{Sec:PathIntegralSubregionSubdomain} provides the necessary definitions for applying the singularity-suppression mechanism to static spherically symmetric spacetimes. Section~\ref{Sec:ActionPhaseFactor} analyzes the effect of higher-order curvature invariants on the oscillatory integrand in the path integral for two exemplary families of static spherically symmetric spacetimes. We conclude and finish with a discussion in Section~\ref{Sec:Discussion}.

\section{Singularity-suppression selection-principle for the dynamics of gravity}\label{Sec:DynamicalSingularitySuppression}

The idea of suppression of singular spacetimes in the Lorentzian gravitational path integral can be motivated from the finite-action principle by Barrow and Tipler~\cite{Barrow:1988gzc,Barrow:2019gzc} and its profound consequences for cosmology of the early universe~\cite{Barrow:2019gzc}. Grounded on the fundamental role of the action as a starting point for deriving the dynamics and in retaining all gauge symmetries, the finite-action principle proposes the value of the action to serve as a guidance for identifying physically realistic models of the universe.
More precisely, the action of the universe integrated over its entire past and future history should be finite~\cite{Barrow:1988gzc}. \footnote{The finite-action principle can be modified to demand finiteness of quantum transition amplitudes from the early universe to present field configurations~\cite{Jonas:2021xkx}. A principle of finite amplitudes appears to compile well with the Hartle-Hawking no-boundary proposal~\cite{Hawking:1981gb,Hartle:1983ai}, see~\cite{Jonas:2021xkx}. }
 
Demanding finiteness of the action enforces the universe to have a finite spatial volume and lifetime, and significantly constrains modified theories of gravity~\cite{Barrow:1988gzc}. In the context of quadratic gravity~\cite{Salvio:2018crh}, the requirement of a finite cosmological action suppresses anisotropies and inhomogeneities in the early universe and favors inflationary initial conditions~\cite{Barrow:2019gzc,Lehners:2019ibe,Lehners:2023fud}. Similarly, in Ho$\check{\text{r}}$ava-Lifshitz gravity~\cite{Wang:2017brl} the finite-action principle imposes the early universe to be flat and homogenous~\cite{Chojnacki:2021ves}.

Altogether, given an action, the finite-action principle suggests to identify physically viable solutions based on a finite on-shell value of the action.

From a reversed viewpoint, given a collection of physically unrealistic configurations, such as spacetimes with a curvature singularity, the requirement of divergence of the action evaluated on these configurations may serve as a directed method to construct actions in situations where the fundamental theory is unknown, such as in quantum gravity. Importantly, the singularity-suppression selection-principle motivated in this section demands that the destructive interference of singular spacetimes is caused solely by the purely gravitational part of the action, i.e., independently of interactions between the gravitational field and other matter fields relevant at low energies, such as the electromagnetic field.
The unphysical configurations with a spacetime curvature singularity, generically, will be off-shell with respect to an action the above divergence property. The motivation to search for actions with the above property is as follows: Whenever the action $S$ in the path integral~\eqref{eq:PathIntegralFormal} attains a large absolute value for a given spacetime with a curvature singularity and neighboring spacetimes in configuration space~\footnote{The definition of ``neighboring" configurations in a full path integral for gravity is subtle, however, in the examples considered in Sections~\ref{Sec:PathIntegralSubregionSubdomain} and~\ref{Sec:ActionPhaseFactor}, the effective configuration spaces are finite-dimenional and characterized by one or two real parameters which the determine the regularity of polynomial curvature invariants for these spacetimes. In these toy-model examples, by a neighborhood of a given spacetime configuration we mean an open interval around the point given by the values of these parameters for this spacetime.}, the rapidly oscillating phase factor is expected to lead to the destructive interference between these configurations and therefore to their suppression in the path integral, i.e.,
\be\label{eq:SuppressionMechanism}
\eval{\abs{S}}_{\qty{g_{\mu\nu}^\text{singular}}} \gg 1 \,\,\,\Rightarrow \,\,\, \qty{g_{\mu\nu}^\text{singular}} \text{ suppressed in } Z\,.
\ee
Let us emphasize that the implication~\eqref{eq:SuppressionMechanism} is a property intrinsic to Lorentzian path integrals. In a Euclidean path integral obtained through a formal Wick rotation, the argument~\eqref{eq:SuppressionMechanism} does not apply. As gravitational actions generically fail to be positive definite~\cite{Gibbons:1977ab,Gibbons:1978ac}, in a Euclidean path integral, spacetimes with a divergent action may be exponentially suppressed or enhanced.

In what follows, concretely, we will demand that the action diverges for spacetimes with a scalar-curvature singularity, in the limit in which the integration is extended up to the singularity. Such a reasoning has been originally applied in the context of black-hole singularity resolution in the Lorentzian gravitational path integral~\cite{Borissova:2020knn}. Thereby it was illustrated that an action quadratic in the curvature and including the Kretschmann scalar $R_{\mu\nu\rho\sigma}R^{\mu\nu\rho\sigma}$ may suffice to suppress the singular Schwarzschild and Kerr black-hole solutions to general relativity~\cite{Borissova:2020knn}. By contrast, the Einstein-Hilbert action and, more generally, any action constructed only from invariants of the Ricci tensor $R_{\mu\nu}$ cannot guarantee the destructive interference of these spacetimes~\cite{Borissova:2020knn}. The latter observation follows from the fact that the Schwarzschild and Kerr spacetimes are vacuum solutions to the Einstein equation, i.e., $R_{\mu\nu}=0$, and thus such an action would evaluate to zero. It should be noted that by including Riemann curvature invariants beyond quadratic order in the action, none of the singular black-hole spacetimes of general relativity is expected to remain a solution to the dynamics derived from such an action, i.e.,~these spacetimes become off-shell. In this sense the suppression of spacetimes with a curvature singularity in the path integral by means of a higher-curvature action should be viewed as part of a suppression mechanism for off-shell singular spacetimes.~\footnote{This work is focused thoroughly on the suppression (via a divergent action) of spacetimes with a curvature singularity, which are {\it not} stationary points of the action. We do not make any preassumptions about the way on-shell spacetimes contribute to the path integral. To the best of our knowledge, the role of on-shell configurations exhibiting a divergent action has not been addressed in explicit evaluations of Lorentzian path integrals for gravity via, e.g., Picard-Lefschetz theory~\cite{Feldbrugge:2017fcc,Feldbrugge:2017kzv}.} One would ultimately hope that on-shell spacetimes representing solutions to the dynamics derived from such a modified action, correspond to regular geometries. Whether this is indeed the case remains a key open question in quantum gravity.

The main objective of this work is to extend the observations in~\cite{Borissova:2020knn} by demanding that {\it all} off-shell spacetimes with a curvature singularity are suppressed in the Lorentzian path integral through a divergent action which causes destructive interference. We will show that such a requirement on the action appearing in the path integral can not be satisfied by local corrections of the form $\mathcal{R}^n$ or $\mathcal{R}\Box^k \mathcal{R}$ for any finite $n,k\in \mathbb{N}$. 
To that end, in the next Sections~\ref{Sec:PathIntegralSubregionSubdomain} and~\ref{Sec:ActionPhaseFactor} we analyze the phase factor of the path integral~\eqref{eq:PathIntegralFormal} as determined by the spacetime integral over these invariants, for different examples in a static spherically symmetric context.

\section{Path integral over subdomain of geometries}\label{Sec:PathIntegralSubregionSubdomain}

In the formal expression for the gravitational path integral~\eqref{eq:PathIntegralFormal}, all possible spacetimes modulo diffeomorphisms are integrated over. The integration domain contains in particular the subset of diffeomorphism-inequivalent static spherically symmetric spacetimes, i.e.,
\be\label{eq:DomainRestriction}
\{[g_{\mu\nu}]\} \supset \{[ \,g_{\mu\nu} \text{ static spherically symmetric}]\}\,,
\ee
whose line element in spherical coordinates can be written as
\be\label{eq:MetricSphericalSymmetry}
\dd{s^2} = -A(r)\dd{t}^2 + B(r)\dd{r}^2 + r^2   \dd{\Omega}^2\,.
\ee
Here $\dd{\Omega}^2 =  \dd{\theta}^2+ \sin^2(\theta) \dd{\phi}^2$ is the area element on the unit $2$-sphere $\mathcal{S}^2$. In what follows $(t,r,\theta,\phi)$ denote dimensionless coordinates, whereby $t$ and $r$ are related to their dimensionful versions through a rescaling by the Planck time and Planck length, respectively.

Later, in Section~\ref{Sec:ActionPhaseFactor}, the singularity-suppression mechanism is studied on the example of two separate families of static spherically symmetric spacetimes. Each of these families should be viewed as representing a proper subset of the integration domain over static spherically symmetric geometries~\eqref{eq:DomainRestriction}. For the argument presented in this paper it is immaterial whether these families arise as a first-order approximation or rather represent an exact expression of some geometry. Mostly we shall refrain from regarding the presented examples as approximate expressions, as our interest lies in the specific form of their curvature invariants. Generally, there is no reason to exclude geometries of precisely the form analyzed in Section~\ref{Sec:ActionPhaseFactor} in the gravitational partition function, since the latter is assumed to integrate over equivalence classes of any four-dimensional spacetime metrics. The fact that some of the considered spacetimes are not asymptotically flat as $r\to \infty$ is not of importance either, as the focus of our discussion will be the gravitational partition function within a Planckian spacetime volume around the potential singularity at $r=0$. It may be possible to combine geometries appearing in this regime, which would not be asymptotically flat at infinity, with spacetimes at larger values of the radial coordinate, which are asymptotically flat, and thereby obtain an effective description of spacetime valid at all scales. The task of deriving such an effective geometry which interpolates between macroscopic and microscopic scales, if it exists, is one of the main challenges in quantum gravity and not addressed in this work.

Altogether, we aim to analyze the effect of different types of local corrections to the action $S$ entering the path integral~\eqref{eq:PathIntegralFormal}, with regard to the requirement of dynamically suppressing all spacetimes with a curvature singularity out of a given family of static spherically symmetric spacetimes. The latter requirement concerns the structure of the gravitational action in the high-energy regime where quantum-gravitational effects, e.g., reflected in higher-curvature contributions, become significant. For this reason we consider the spacetime region representing a thin spherical shell of Planckian size around the origin $r=0$ of the radial coordinate, which marks the location of the potential curvature singularity of all spacetimes studied later. The coordinate $r$ is allowed to range in $ r\in  (\epsilon , 1)$ where $\epsilon>0$ represents a dimensionless UV cutoff regulator. For the proposed singularity-suppression condition, the divergence of the action for singular spacetimes ought to originate intrinsically from the integration over the radial coordinate, depending on which invariants are used to construct the Lagrangian density. The integral over the time coordinate, on the other hand, contributes only by a constant factor which is irrelevant for the dynamical singularity suppression mechanism. We will therefore cut off the time integral and choose a regularized finite time interval of unit length. Consequently, the action for static spherically symmetric spacetimes~\eqref{eq:MetricSphericalSymmetry} is written as
\be\label{eq:ActionSpherical SymmetryPowerLaw}
S_{\epsilon} \equiv \int \dd[4]{x}\sqrt{-g} \mathcal{L} =4\pi  \sum_i \gamma_i \int_{\epsilon}^{1}\dd{r}  \sqrt{-g_r(r)} \mathcal{K}_i(r) \,.
\ee
Here $\mathcal{L}$ represents an a priori unspecified covariant Lagrangian density constructed from polynomial invariants $\mathcal{K}_i$ of the Riemann curvature tensor $\mathcal{R}$ and covariant derivatives $\nabla$ of the metric, with couplings $\gamma_i$. The $r$-dependent part of the metric determinant is denoted by $g_r(r)=-r^4 A(r)B(r)$ and we have integrated over the time and angle coordinates. 

Of ultimate interest for the question of destructive interference of spacetimes with a curvature singularity, is the value of the action~\eqref{eq:ActionSpherical SymmetryPowerLaw} in the limit in which the lower integration boundary $\epsilon$ approaches zero. According to~\eqref{eq:SuppressionMechanism}, the suppression of a set of neighboring off-shell singular spacetimes can be achieved if the value of the action for these spacetimes is infinite. In  Section~\ref{Sec:ActionPhaseFactor} we analyze the value of spacetime integrals built from local higher-curvature and higher-derivative corrections in the limit $\epsilon\to 0$ for exemplary families of static spherically symmetric spacetimes. The analysis is based on polynomial invariants of the Riemann tensor and covariant derivatives of the metric which can be written as
\be
\mathcal{R}^n,\, \mathcal{R}\Box^k\mathcal{R}\,,\,\,\, n,k \in \mathbb{N}\,.
\ee

For future reference we note that for static spherically symmetric spacetimes written in the form~\eqref{eq:MetricSphericalSymmetry}, the action of the d'Alembert operator on functions $f(r)$ is given by
\be\label{eq:BoxAction}
\Box f = \frac{1}{B} \qty[\qty(\frac{A'}{2A}- \frac{B'}{2B}+\frac{2}{r})f'+f'']\,, 
\ee
where a prime denotes differentiation with respect to $r$.
To determine which ranges out of a given set of singular spacetimes can be suppressed in the path integral through a divergent action, we will moreover repeatedly use that
\be\label{eq:IntegralFormula}
\lim_{\epsilon \to 0} \int_{\epsilon}^1 \dd{r} r^\gamma < \infty \, \Leftrightarrow \, \gamma >-1\,.
\ee

Finally, let us take a moment to comment on equation~\eqref{eq:ActionSpherical SymmetryPowerLaw}, which represents only the bulk contribution to the full action. Generically, in the presence of boundaries, both the bulk and the boundary contributions are crucial for determining the value of the action.~\footnote{More generally, topological terms also contribute to the value of the action. Here we assume that this contribution is finite and therefore irrelevant to the discussion of necessary conditions for divergence of the action evaluated on singular spacetimes.} In our case the order in curvatures and specific form of the action are not fixed a priori, and thus the structure of the boundary term which would be necessary for a well-defined variational principle is not clear. For spherically symmetric spacetimes~\eqref{eq:MetricSphericalSymmetry}, the  boundary action will contain an integral over the radial coordinate of the radial part of the square root of the determinant of the induced metric at the boundary, times a combination of curvature invariants. As the determinant of the spatial metric contributes only by a power law in $r$ with constant exponent, the leading-order behavior at $r=0$ of the boundary integral is impacted by which curvature invariants are included in the boundary action. Such an interplay between the square root of the spatial determinant and the leading-order power law of a given curvature invariant in the boundary action, is of the same type of interplay as the one between the square root of the full metric determinant and a curvature invariant in the bulk action. Both integrals take the form~\eqref{eq:IntegralFormula} for some exponents $\gamma$ and $\gamma'$. Thus, in order to derive necessary conditions in form of which types of curvature invariants can achieve the divergence of an action integral, it is sufficient to focus on the bulk action integral~\eqref{eq:ActionDivergenceCondition}.~\footnote{There might exist special cases in which separately divergent bulk and boundary actions of some singular spacetimes cancel each other out, resulting in a zero value of the full action for these spacetimes. At this stage, we can not exclude that such a situation occcurs. However, as we shall see, the results of Section~\ref{Sec:ActionPhaseFactor} support generic non-perturbative actions in order to satisfy the singularity-suppression principle. The leading-order terms in the bulk and boundary contribution to such an action would have to be related in a very specific way in order to have this type of cancellation happening. We do not expect such a cancellation to occur generically.
}

\section{$\imath \mathbf{S}\qty[ \mathcal{R}^n, \mathcal{R}\Box^k\mathcal{R}]$  phase factors}\label{Sec:ActionPhaseFactor}

The oscillatory behaviour of the imaginary phase factor~$\imath S$ in the path integral~\eqref{eq:PathIntegralFormal} is dictated by the value of the action evaluated on the different geometries in configuration space. In the following the value of spacetime integrals constructed from local invariants $\mathcal{R}^n$ and $\mathcal{R}\Box^k \mathcal{R}$ is studied for two exemplary families of static spherically symmetric spacetimes.

\subsection{Example I\,: $A(r)=B(r)^{-1}=1-r^\alpha$}\label{SecSub:Example1}

As a first example we consider a subset of static spherically symmetric spacetimes~\eqref{eq:MetricSphericalSymmetry} characterized by $A(r)=B(r)^{-1}$, where $A$ is given by
\be\label{eq:ExampleIFunctionA}
A(r) = 1-  r^\alpha\,,
\ee
with arbitrary real exponents $\alpha \in \mathbb{R}$. In~\eqref{eq:ExampleIFunctionA} a potential free non-vanishing coefficient in front of $r^\alpha$ has been fixed to one. It is possible to show that such a coefficient appears in all $\mathcal{R}^n$ invariants for these spacetimes only as a global factor which counts the number of curvatures, cf.~Appendix~\ref{SecSub:ZMInvariantsExampleI}. Therefore the value of this coefficient will be irrelevant for the discussion of spacetime integrals built from these invariants.
Spacetimes with lapse function of the form~\eqref{eq:ExampleIFunctionA} include the singular Schwarzschild spacetime for $\alpha = -1$, whereas the value $\alpha = 2$ describes a spacetime with finite constant curvature. In this case~\eqref{eq:ExampleIFunctionA} can be regarded as the next-to-leading order expansion around $r=0$ of various types of non-singular black holes~\cite{Frolov:2016pav} with smooth de Sitter cores, such as the Bardeen~\cite{Bardeen:1968bh},  Dymnikova~\cite{Dymnikova:1992ux} and Hayward~\cite{Hayward:2005gi} black-hole spacetimes. These, and spacetimes which to first order at $r=0$ are described by other values of $\alpha$, can arise, e.g., as the result of renormalization-group (RG) improvement inspired by the asymptotic-safety scenario for quantum gravity~\cite{Bonanno:2000ep,Falls:2010he,Torres:2014gta,Adeifeoba:2018ydh,Pawlowski:2018swz,Platania:2019kyx,Platania:2023srt,Torres:2017ygl,Borissova:2022mgd,Kofinas:2015sna,Cai:2010zh,Bonanno:1998ye,Emoto:2005te}, or as effective geometries in loop quantum gravity~\cite{Rovelli:2014cta,Han:2023wxg}. As previously anticipated, for our purposes we shall refrain from viewing~\eqref{eq:ExampleIFunctionA} as an approximate expression. Our main motivation for choosing the specific family of spacetimes with lapse function given by~\eqref{eq:ExampleIFunctionA}, is that it features two uncountable sets of singular and regular spacetimes depending on the value of the parameter $\alpha$. Specifically, we will see that all non-vanishing Riemann polynomial invariants $\mathcal{R}^n$ in this case have a particularly simple structure and are uniquely determined by the value of the parameter $\alpha$. The latter can therefore be thought of as labelling physically distinct configurations in a ``mini-superspace" path-integral model over this subfamily of geometries. Despite the square root of the radial part of the metric determinant of the spacetimes~\eqref{eq:ExampleIFunctionA} being parameter-independent, $\sqrt{-g_r(r)} = r^2$, it turns out that no invariant $\mathcal{R}^n$ for finite $n\in \mathbb{N}$ can lead to a divergent spacetime integral for the singular spacetimes on approach to the singularity, thereby illustrating the non-triviality of the requirement of dynamical singularity suppression.

The Kretschmann curvature scalar for the family of spacetimes~\eqref{eq:ExampleIFunctionA} takes the form
\be\label{eq:KretschmannScalar}
R_{\mu\nu\rho\sigma}R^{\mu\nu\rho\sigma} \propto r^{2\qty(\alpha - 2)}\,,
\ee
where the proportionality coefficient is a polynomial of order four in $\alpha$ with no real zeros. Thus, all spacetimes with
\be
\alpha <2
\ee
exhibit a curvature singularity as $r\to0$ and should be suppressed in the gravitational path integral. \\

To achieve the suppression of off-shell geometries with a scalar-curvature singularity in the path integral, according to the mechanism advocated in Section~\ref{Sec:DynamicalSingularitySuppression}, the action~\eqref{eq:ActionSpherical SymmetryPowerLaw} should evaluate to positive or negative infinity for singular spacetimes, as the integration is extended up to the singularity. Concretely, for the family of static spherically symmetric geometries described by \eqref{eq:ExampleIFunctionA}, the value of the action~\eqref{eq:ActionSpherical SymmetryPowerLaw} should be infinite for all spacetimes with $\alpha <2$, in order to guarantee their destructive interference. 

Let us initially assume that the action input in the path integral~\eqref{eq:PathIntegralFormal} is given by the classical Einstein-Hilbert action with Ricci scalar $R$. For the spacetimes~\eqref{eq:ExampleIFunctionA} the latter is given by
\be\label{eq:RicciScalar}
R(r) \propto r^{\alpha-2}\,.
\ee
Here we have dropped the proportionality factor representing a quadratic polynomial in $\alpha$. The latter vanishes, e.g., for $\alpha = -1$, corresponding to the Ricci-flat Schwarzschild vacuum solution to Einstein's field equations. In what follows we will not be concerned with possible real roots of polynomials in the free parameter $\alpha$, which appear in the monomials with different powers of $r$ in a given curvature invariant. To investigate the destructive interference between neighboring singular spacetimes, associated with real intervals of the parameter $\alpha$, only the powers of $r$ which can possibly appear in a given curvature invariant are of interest. We will proceed in the same way for the second example later in Section~\ref{SecSub:Example2}.

The action integral~\eqref{eq:ActionSpherical SymmetryPowerLaw} with Ricci scalar~\eqref{eq:RicciScalar} becomes
\be
{S_{\epsilon}}_{\text{EH}} = 4\pi  \qty( \frac{1}{2\kappa }) \int_{\epsilon}^{1} \dd{r} r^2 R(r)  \propto  \int_{\epsilon}^{1} \dd{r} r^\alpha \,.
\ee
By means of the integral identity~\eqref{eq:IntegralFormula}, the latter expression is finite as $\epsilon\to 0$, provided $\alpha >-1$. Consequently, the Einstein-Hilbert action evaluated for the singular spacetimes in the parameter range $\alpha \in (-1,2)$ is finite. On these grounds, the Einstein-Hilbert action is not suitable to achieve destructive interference in the path integral between all the singular configurations contained in this example set of spacetimes. At first sight, one might attribute this result to the fact that the Ricci scalar contains only two derivatives of the metric and therefore could simply be not singular enough to render the action integral divergent. As the square root of the metric determinant for the spacetimes~\eqref{eq:ExampleIFunctionA} is independent of the parameter $\alpha$, one would expect that there exists a finite number $n\in \mathbb{N}$ of contracted Riemann curvature tensors, $\mathcal{R}^n$, beyond which the action becomes divergent for singular spacetimes on approach to the singularity. This turns out not to be the case due to the specific form of such invariants evaluated for the metrics~\eqref{eq:MetricSphericalSymmetry}, as we shall illustrate next.

 Remarkably, it is possible to show that for any given representative labeled by $\alpha$ out of the family~\eqref{eq:ExampleIFunctionA}, all curvature invariants corresponding to powers of the Riemann tensor can be polynomially generated by a single one. This can be seen, e.g., by computing the Zakhary-McIntosh (ZM) invariants~\cite{Carminati:1991zm,Mcintosh1997:zm,Carminati:2002zm} which form a polynomial basis for the set $\{\mathcal{R}^n\,|\,n\in \mathbb{N}\}$ of an arbitrary four-dimensional spacetime. In Appendix~\ref{SecSub:ZMInvariantsExampleI} we show that for the spacetimes~\eqref{eq:ExampleIFunctionA}, any polynomial invariant built from $n\in \mathbb{N}$ contracted Riemann curvature tensors can be written as
\be
\mathcal{R}^n \propto r^{n(\alpha-2)}\,.
\ee
It thereby follows that the contribution $s_\epsilon\qty[\mathcal{R}^n]$ of an invariant $\mathcal{R}^n$ to the full action~\eqref{eq:ActionSpherical SymmetryPowerLaw} is given by
\be\label{eq:Sn}
s_\epsilon\qty[\mathcal{R}^n] \propto \int_{\epsilon}^{1} \dd{r} r^2 r^{n(\alpha -2)} \,.
\ee
The latter integral converges in the limit $\epsilon\to 0$, using equation~\eqref{eq:IntegralFormula}, for $\alpha > 2-3/n$. Accordingly, such a term in the action diverges in the limit $\epsilon\to 0$ for spacetimes with a parameter value
\be\label{eq:ActionDivergenceCondition} 
\alpha \leq 2 - \frac{3}{n}\,.
\ee
As a consequence, for any finite maximal number $n\in \mathbb{N}$ of contracted Riemann tensors $\mathcal{R}^n$ used to form an action $S$, there remains a range of singular spacetimes described by parameter values $\alpha \in (2-3/n,2)$ whose suppression by means of destructive interference in the Lorentzian path integral~\eqref{eq:PathIntegralFormal} is not guaranteed. 

\begin{figure}[!t]
	\includegraphics[width=\linewidth]
	{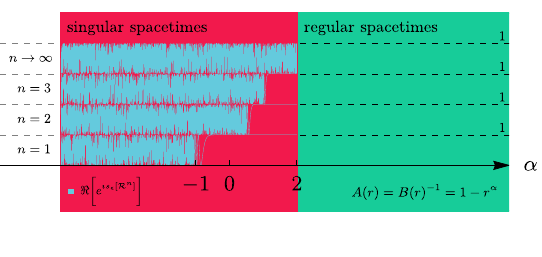}
	\caption{Illustration of the dynamical suppression of singular spacetimes in the Lorentzian gravitational path integral via terms $\mathcal{R}^n$ with $n\in \mathbb{N}$ in the action for the one-parameter family of static spherically symmetric spacetimes~\eqref{eq:ExampleIFunctionA}. Spacetimes with $\alpha < 2$ exhibit a scalar-curvature singularity at $r= 0$. Their suppression via destructive interference originates from a divergent action if the integration is extended up to the singularity, as represented by the blue blocks. These show the real part of the highly oscillatory path-integrand factors $e^{\imath s_\epsilon\qty[\mathcal{R}^n]}$ where $s_\epsilon\qty[\mathcal{R}^n]$ is given by~\eqref{eq:Sn} and the UV cutoff regulator has been set to $\epsilon = 10^{-4}$. The complete dynamical singularity suppression in the path integral is achieved only in the limit $n\to \infty$, corresponding to an action of infinite order in the curvature. }
	\label{Fig:IntegrandSingularSuppression}
\end{figure}

Figure~\ref{Fig:IntegrandSingularSuppression} shows the path-integrand factors $e^{\imath s_\epsilon\qty[\mathcal{R}^n]}$ for different $n\in \mathbb{N}$. The oscillatory patterns of the real (and imaginary) part illustrate the destructive interference between spacetimes with corresponding values of $\alpha$, caused by the divergence of the action terms $s_\epsilon\qty[\mathcal{R}^n]$ in the limit $\epsilon\to 0$. Higher-curvature invariants push the bound of suppressed singular spacetimes to increasing values of $\alpha$. The value $\alpha =2$ which marks the border between spacetimes with and without a curvature singularity is reached only asymptotically for $n\to \infty$, i.e., in the limit in which terms with infinitely many contracted curvature tensors are included in the action.

The previous observations allow us to extract a central result for the full gravitational Lorentzian path integral~\eqref{eq:PathIntegralFormal}: 
In order to suppress all off-shell spacetimes with a scalar-curvature singularity through a divergent value of the action, bare actions consisting of only finitely many contracted curvature tensors, $\mathcal{R}^n$ for finite $n\in \mathbb{N}$, are insufficient. 

On the basis of the previous results, one may ask whether curvature-derivative invariants, i.e., polynomial Riemann invariants including explicit covariant derivatives of the metric, might achieve the suppression of the singular spacetimes from the family~\eqref{eq:ExampleIFunctionA} in the path integral. In fact, it is straightforward to verify that invariants with two insertions of the covariant d'Alembert operator, evaluated on the spacetimes~\eqref{eq:ExampleIFunctionA}, contain the monomials
\be\label{eq:RBox2RExample1}
\mathcal{R}\Box^2 \mathcal{R}\,\, \ni \,\,r^{-8 + z\alpha}\,,\,\,\, z=2,3,4 \,.
\ee
Combining the first term with the factor $r^2$ originating from the square root of the metric determinant, using~\eqref{eq:IntegralFormula},  the corresponding spacetime integral diverges as $\epsilon\to 0$ if $\alpha\leq 5/2$. Therefore the curvature-derivative invariants~\eqref{eq:RBox2RExample1} are indeed sufficient to suppress all singular spacetimes from the family~\eqref{eq:ExampleIFunctionA} in the path integral. \footnote{Certain polynomial curvature-derivative invariants of the form $\mathcal{R}\Box^k \mathcal{R}$ with $k\in \mathbb{N}$ may also act as black-hole horizon-detecting functions~\cite{Page:2015aia, McNutt:2017gjg,McNutt:2017paq,Coley:2017vxb}.} In particular, this result strengthens the aforementioned expectation that, due to the parameter-independent metric determinant given by a finite constant power of $r$, there should exist invariants with a finite number of metric derivatives, for which the spacetime integral diverges for all singular spacetimes upon integrating up to the singularity. The insufficiency of invariants $\mathcal{R}^n$ for finite $n\in \mathbb{N}$ to achieve the destructive interference between the singular spacetimes in the family~\eqref{eq:ExampleIFunctionA}, arises merely due to non-trivial cancellations of monomials which could in principle appear in these invariants. For instance, replacing~\eqref{eq:ExampleIFunctionA} by $A(r)=C-r^\alpha$ with a constant $C\neq 1$, generates an $\alpha$-independent term $\propto (C-1)^2 r^{-4}$ in the Kretschmann scalar $R_{\mu\nu\rho\sigma}R^{\mu\nu\rho\sigma}$. The latter in this case is therefore singular for all values of $\alpha$. As the metric determinant remains unchanged, $\sqrt{-g_r(r)}=r^2$, the Kretschmann scalar for this modified family of spacetimes would already render the action divergent in the limit $\epsilon\to 0$ and thereby suppress all of these singular geometries in the path integral.

To investigate more generally the properties of local invariants $\mathcal{R}^n$ and $\mathcal{R}\, \Box^k  \mathcal{R}$ for $n,k\in \mathbb{N}$ as singularity-suppression invariants in the action, in the next section we analyze a second example of static spherically symmetric metrics.

\subsection{Example II\,: $A(r)=r^\alpha,\, B(r)= r^\beta$}\label{SecSub:Example2}

As a second example we consider static spherically symmetric spacetimes~\eqref{eq:MetricSphericalSymmetry} where the two free functions are described by a power law
\be\label{eq:ExampleIIFunctionsAB}
A(r) =  r^\alpha\,\,\, \text{and}\,\,\, B(r)= r^\beta\,
\ee
with arbitrary real exponents $\alpha, \beta \in \mathbb{R}$. When viewed as a leading-order expansion around $r=0$, e.g., quadratic functions, for which $\alpha=\beta =2$, appear notably as one family of solutions to quadratic gravity in static spherically symmetric settings~\cite{Stelle:1977ry,Holdom:2002xy,Holdom:2016nek,Holdom:2019ouz}. Other values of the parameters $\alpha,\beta$ can describe other families of solutions to quadratic gravity~\cite{Stelle:1977ry,Holdom:2002xy,Lu:2015psa} and also appear in the application of RG-improvement of gravitational actions~\cite{Borissova:2022jqj}. 

In~\eqref{eq:ExampleIIFunctionsAB} we have used the freedom of rescaling the time coordinate to eliminate the coefficient in front of $r^\alpha$ and have additionally set the coefficient in front of $r^\beta$ to one for convenience. The resulting Kretschmann scalar is given by
\be
R_{\mu\nu\rho\sigma}R^{\mu\nu\rho\sigma} = 4 r^{-4} - 8 r^{-4-\beta}+  p(\alpha,\beta)r^{-4-2\beta}
\ee
with a lengthy polynomial $p(\alpha,\beta)$ which we refrain from displaying here.
It is straightforward to verify that the Kretschmann scalar is singular unless $\alpha=\beta = 0$, corresponding to the flat Minkowski metric. Accordingly, we demand that all spacetimes described by parameter values $(\alpha,\beta)\in \mathbb{R}^2{\setminus}\{(0,0)\}$ are suppressed in the gravitational path integral. A primary motivation for choosing this example is the following: 
While the degree of regularity of the spacetimes~\eqref{eq:ExampleIIFunctionsAB}, as measured by the exponents of the monomials in $r$ appearing in curvature invariants, depends only on $\beta$, the parameter $\alpha$ enters the square root of the metric determinant,
\be\label{eq:MetricDeterminantExampleII}
\sqrt{-g_r(r)} = r^{2+\frac{\alpha}{2}+\frac{\beta}{2}}\,.
\ee
Setting $\alpha\neq 0, \beta =0$ as a first subcase allows us to state the general impossibility of complete dynamical singularity resolution via local terms of the form $\mathcal{R}^n$ and $\mathcal{R}\Box^k \mathcal{R}$ for finite $n,k \in \mathbb{N}$. Subsequently, we consider the case $\alpha=0,\beta\neq 0$, by which the effects and power of both types of local curvature invariants as singularity-suppression invariants in the action can be distinguished. Finally, we analyze the situation for general $\alpha,\beta\neq0$ and reconfirm the previous observations. Since for our discussion only the global structure of an invariant matters for determining which monomials in $r$ can appear, we will focus only on the invariants $R^n$ and $R\Box^k R$, where $R$ denotes the Ricci scalar. We shall first keep $n,k \in \mathbb{N}$ arbitrary and unrelated. Occasionally, we will set $n=k+2$ to allow for a comparison of the effects of both types of invariants when an equal number of metric derivatives is involved.

\subsubsection{Case I\,: $\alpha\neq 0,\,\beta =0 $}\label{SecSubSub:Example2Case1}

\begin{figure*}[!t]
	\centering
	\includegraphics[width=1.73in]{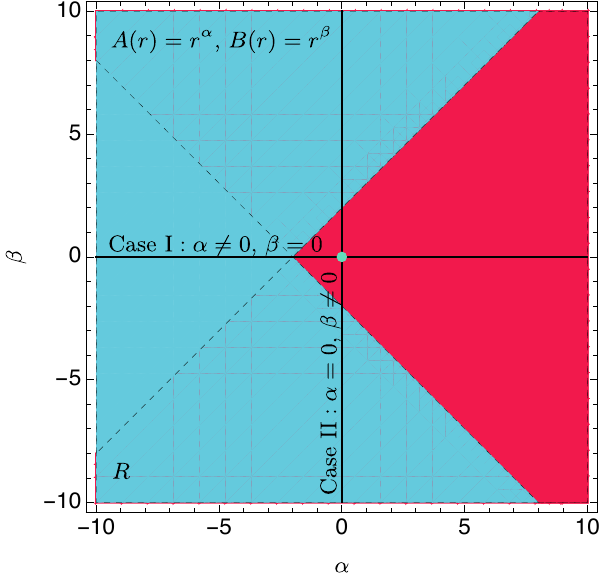}
	\includegraphics[width=1.73in]{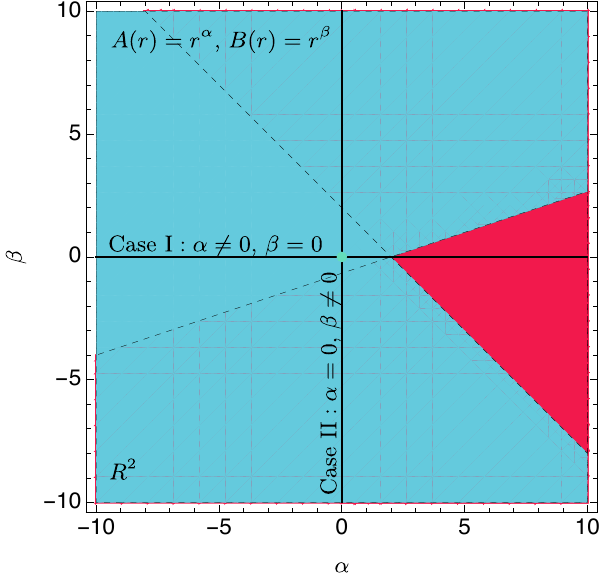}
	\includegraphics[width=1.73in]{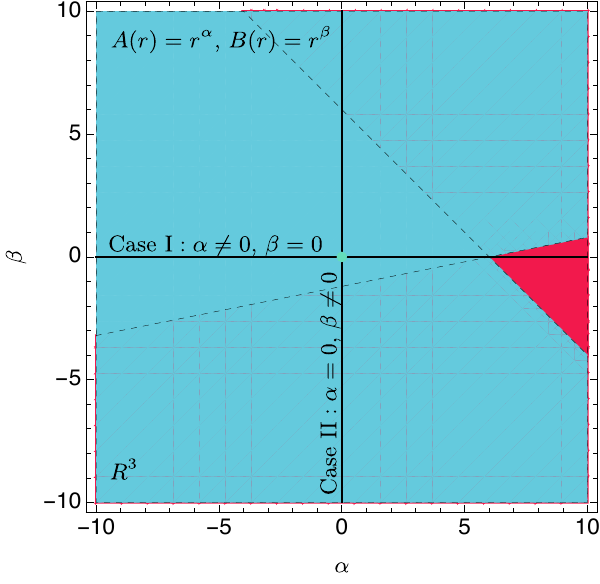}
	\includegraphics[width=1.73in]{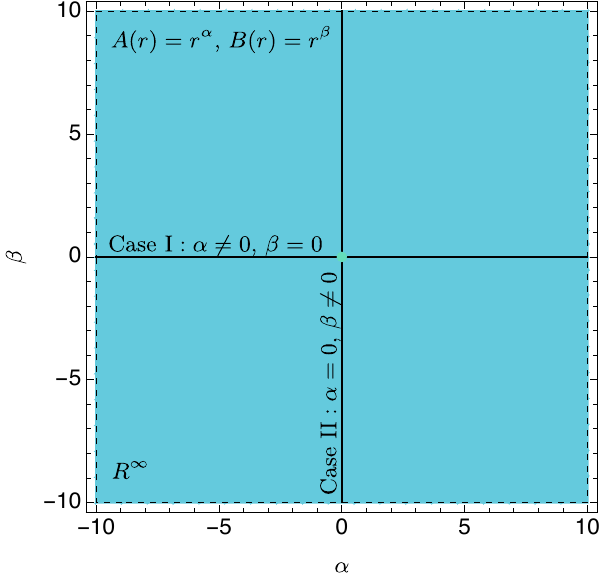}
	\caption{Parameter space $\{(\alpha,\beta)\} =  \mathbb{R}^2\setminus{\{(0,0)\}}$ of singular spacetimes with power-law metric functions~\eqref{eq:ExampleIIFunctionsAB} and scalar-curvature singularity at $r=0$, subdivided into two regions: singular spacetimes which {\it can} be suppressed in the Lorentzian gravitational path integral via terms $R^n$ for $n=1,2,3,\infty$ (four panels, from left to right) in the action (blue region), and spacetimes for which the covariant spacetime integral over these invariants remains finite on approach to the singularity and thus the destructive interference of these spacetimes through a rapidly oscillating path-integral phase factor {\it can not} be achieved (red region). The green dot at $(0,0)$ labels the non-singular flat Minkowski spacetime and is not part of the singular parameter space. The suppression of all case-I singular spacetimes (cf.~Section~\ref{SecSubSub:Example2Case1}) along the horizontal axis $\{\beta=0\}$ can only be achieved by $R^n$ for $n\to \infty$. For the suppression of all case-II singular spacetimes (cf.~Section~\ref{SecSubSub:Example2Case2}) along the vertical axis $\{\alpha=0\}$, already the inclusion of quadratic-curvature terms $R^2$ is sufficient. In the general case III, $\{\alpha,\beta\neq 0\}$, (cf.~Section~\ref{SecSubSub:Example2Case3}), an unsuppressed red wedge of singular spacetimes remains unless the action includes terms of infinite order in the curvature, i.e., $R^n$ for $n\to \infty$. }\label{Fig:RNSuppression}
\end{figure*}

\begin{figure*}[!t]
	\centering
	\includegraphics[width=1.73in]{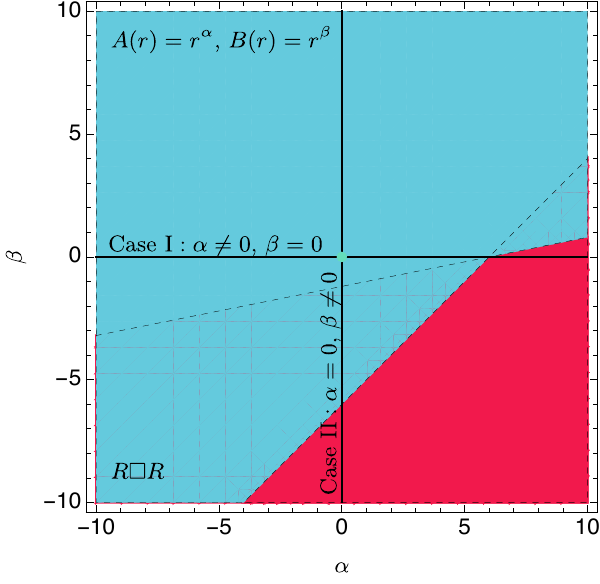}
	\includegraphics[width=1.73in]{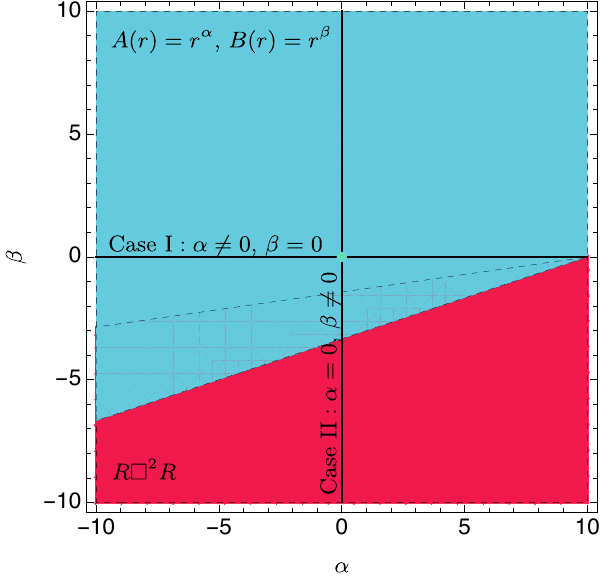}
    \includegraphics[width=1.73in]{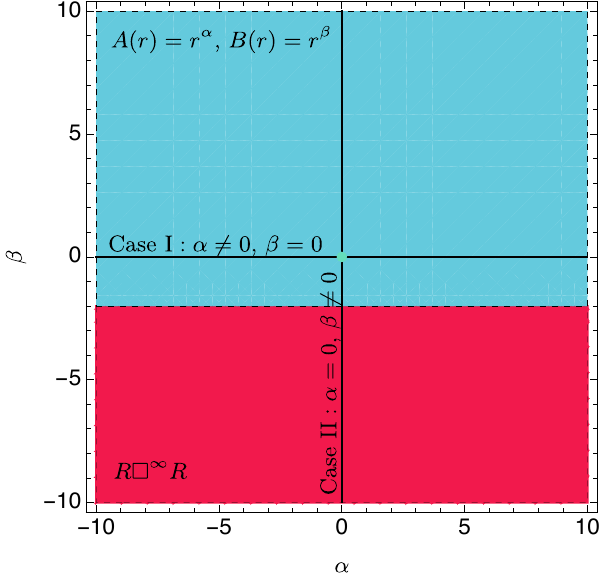}	\includegraphics[width=1.73in]{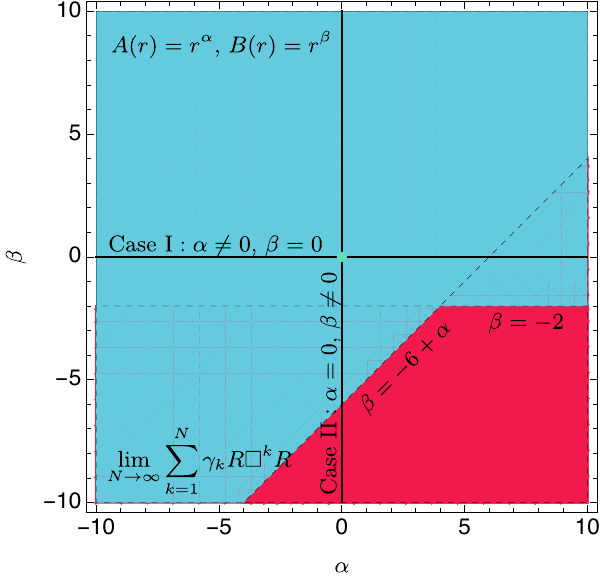}
	\caption{Parameter space $\{(\alpha,\beta)\} =  \mathbb{R}^2\setminus{\{(0,0)\}}$ of singular spacetimes with power-law metric functions~\eqref{eq:ExampleIIFunctionsAB} and scalar-curvature singularity at $r=0$, subdivided into two regions: singular spacetimes which {\it can} be suppressed in the Lorentzian gravitational path integral via terms $R\Box^k R$ for $k=1,2,\infty$ (first three panels, from left to right) in the action (blue region), and spacetimes for which the covariant spacetime integral over these invariants remains finite on approach to the singularity and thus the destructive interference of these spacetimes through a rapidly oscillating path-integral phase factor {\it can not} be achieved (red region). The green dot at $(0,0)$ labels the non-singular flat Minkowski spacetime and is not part of the singular parameter space. The suppression of all case-I singular spacetimes (cf.~Section~\ref{SecSubSub:Example2Case1}) along the horizontal axis $\{\beta=0\}$ can only be achieved by $R\Box^k R$ for $k\to \infty$. Invariants $R\Box^k R$ can not achieve the suppression of all case-II singular spacetimes (cf.~Section~\ref{SecSubSub:Example2Case2}) along the vertical axis $\{\alpha=0\}$. For $k=1$ a maximal range $\beta \in (-6,\infty)$ of case-II spacetimes is suppressed, whereas an increasing number $k$ of insertions of the covariant d'Alembert operator between two curvatures reduces the range of suppressed spacetimes along the $\{\alpha=0\}$ axis to a limit interval $\beta \in (-2,\infty)$ for $k\to \infty$. This result originates from the regularizing effect of the covariant d'Alembert operator~\eqref{eq:BoxAction} acting on homogenous functions when $\beta > -2$. In the general case III, $\{\alpha,\beta\neq 0\}$, (cf.~Section~\ref{SecSubSub:Example2Case3}), the infinite sum $\lim_{N\to \infty}\sum_{k=1}^N \gamma_k R\Box^k R$ in the action with $\gamma_{1,N} \neq 0$ leaves an unsuppressed red wedge of singular spacetimes bounded above by $\beta= - 2$ and $\beta = -6+\alpha$, as shown in the fourth panel. }	\label{Fig:RBoxNRSuppression}
\end{figure*}

In the case $A(r)=r^\alpha$ and $B(r)=1$, an evaluation of the Ricci scalar and application of the action~\eqref{eq:BoxAction} of the d'Alembert operator on a homogeneous function $f(r)=r^\gamma$, yield~\footnote{By computing the elements of the ZM polynomial basis for Riemann invariants, analogously as done for example I in Appendix~\ref{SecSub:ZMInvariantsExampleI}, it is straightforward to see that for all invariants $\mathcal{R}^n$ it holds that $\mathcal{R}^n \propto r^{-2n}$. }
\ba\label{eq:ScalarsContentExampleIICaseI}
R^n\,\,\propto& r^{-2n}\,,\,\,\,&n\in \mathbb{N}\nn\\
R\Box^k R\,\,\propto& r^{-2(k+2)}\,,\,\,\,&k\in \mathbb{N}\,.
\ea
When multiplied with the radial part of the square root of the metric determinant~\eqref{eq:MetricDeterminantExampleII} with $\beta=0$ to form a spacetime integral, using equation~\eqref{eq:IntegralFormula}, these integral terms respectively can suppress singular spacetimes in the ranges
\ba\label{eq:SuppressionRangesExampleIICaseI}
\alpha\leq &-6+4n\,,\,\,\, &n \in \mathbb{N}\nn\\
\alpha\leq& 2+4k\,,\,\,\, &k \in \mathbb{N}\,.
\ea
The inequalities~\eqref{eq:SuppressionRangesExampleIICaseI} illustrate that there cannot exist a universal local invariant $\mathcal{R}^n$ or $\mathcal{R}\Box^k\mathcal{R}$ for any finite $n,k \in \mathbb{N}$, such that all singular spacetimes are suppressed in the path integral. Instead, in this particular subcase, adding such types of terms to the action for $n \to  \infty$ or $k\to \infty$ both provide possible ways of achieving full singularity suppression. This result can also be seen from Figure~\ref{Fig:RNSuppression} and Figure~\ref{Fig:RBoxNRSuppression}. The blue region represents suppressed singular spacetimes, whereas the red region corresponds to unsuppressed singular spacetimes. The green point at the origin marks the non-singular flat Minkowski metric, for which all curvature invariants vanish. In the case of $\mathcal{R}^n$ shown in Figure~\ref{Fig:RNSuppression}, the red region is bounded by two non-horizontal lines whose intersection point lies at $\alpha = -6 + 4n $, cf.~equation~\eqref{eq:SuppressionRangesExampleIICaseI}, along the $\{\beta=0\}$ axis. Increasing $n\in \mathbb{N}$ amounts to a shift of this point to larger values of $\alpha$. The full range $\{\alpha\in \mathbb{R}\setminus{\{0\}}\}$ can be covered only in the limit $n\to \infty$ when the intersection point is shifted to infinity. Similarly, in Figure~\ref{Fig:RBoxNRSuppression} the red region describes the parameter space of singular spacetimes which can be suppressed via terms $\mathcal{R}\Box^k\mathcal{R}$ in the action. The edge point of the red region lies along the $\{\beta=0\}$ axis at a finite value $\alpha = -6  +4k $ for all finite $k\in \mathbb{N}$. Only in the limit $k\to \infty$ this corner reaches infinity, the edge is smoothed out and the separatrix between the red and blue regions becomes a horizonal line at $\{\beta = -2\}$, as we shall see later.

From the suppressed ranges~\eqref{eq:SuppressionRangesExampleIICaseI} it follows that, for this special subcase, the effects of local terms $\mathcal{R}^n$ and $\mathcal{R}\Box^k \mathcal{R}$ in the action are equivalent for $n=k+2$, i.e., when the number of involved metric derivatives are identical. This effect occurs however only due to the action~\eqref{eq:BoxAction} of the d'Alembert operator on homogenous functions $r^\gamma$, which in the case when $A(r)=r^\alpha$ and $B(r)=1$ amounts to a partial derivative,
\be\label{eq:BoxActionExampleIICaseI}
\Box \,r^\gamma \propto r^{\gamma-2} \propto \partial_r^2 \,r^\gamma\,.
\ee
To contrast more generally the suppression powers of the above two types of invariants, next we consider a second subcase characterized by parameter values $\alpha=0$ and $\beta\neq 0$.

\subsubsection{Case II\,: $\alpha= 0,\,\beta \neq 0 $}\label{SecSubSub:Example2Case2}

In the case $A(r)=1$ and $B(r)=r^\beta $, the Ricci scalar for generic $\beta$ contains the monomials
\be\label{eq:RicciScalarExampleIICaseII}
R \,\,\ni \,\, r^{-2-z\beta},\,\,\,z=0,1\,.
\ee
The action~\eqref{eq:BoxAction} of the covariant d'Alembert operator on a homogenous function $r^\gamma$ is given by
\be\label{eq:BoxActionExampleIICaseII}
\Box \, r^\gamma \propto r^{\gamma - 2-\beta}
\ee
and thus an immediate proof by induction establishes that
\ba\label{eq:ScalarsContentExampleIICaseII}
R^n \,\,&\ni \,\,r^{-2n -  z\beta}\,,\,\, \,z=0,\dots,n\,,\,\,\,&n\in \mathbb{N}\nn\\
R\Box^k R \,\,&\ni \,\, r^{-4-2k -\beta k -z\beta}\,,\, \,\, z=0,1,2\,,\,\,\,&k\in \mathbb{N}\,.
\ea

When combined with the square root of the metric determinant~\eqref{eq:MetricDeterminantExampleII} with $\alpha=0$, using the integral identity~\eqref{eq:IntegralFormula} and taking the minimal and maximal values for $z$, such invariants in the action integral are seen to suppress respectively the following parameter ranges of singular spacetimes, 
\ba\label{eq:SuppressionRangesExampleIICaseII}
\beta \leq -6+4n\,,\,\,\beta& \geq \frac{-6+4n}{1-2n}\,,\,\,\,&n\in \mathbb{N} \nn\\
\beta &\geq \frac{2+4k}{1-2k}\,,\,\,\,&k\in \mathbb{N}\,. \nn\\
\ea
Setting $n=k+2$ allows us to compare the effects of the scalar invariants $R^{k+2}$ and $R\Box^{k} R$ as singularity-suppression invariants with an equal number of metric derivatives. $R\Box^{k}R$ sets a lower bound for suppressed singular spacetimes with non-zero parameter $\beta$. The minimum bound $\beta \geq -6$ is achieved by setting $k=1$, as can be seen by comparing the intersection points between the red and blue regions at the $\{\alpha=0\}$ axis in the first three panels of Figure~\ref{Fig:RBoxNRSuppression}. For increasing number $k \in \mathbb{N}$ of insertions of the box operator between two curvatures, the invariants $R\Box^{k} R$ are able to suppress a decreasing range of values for $\beta$, up to the limit point $\beta = -2$ for $k\to \infty$. Therefore, for increasing $k\in \mathbb{N}$ these invariants become less efficient in achieving the divergence of the action for singular spacetimes along the $\{\alpha=0\}$ axis. In particular, there exists a parameter range $\beta \in (-2,\infty)$ of singular spacetimes with $\alpha=0$, which these invariants are not able to suppress in the path integral. This can be seen by looking at the $\{\alpha=0\}$ axis in the third panel in Figure~\ref{Fig:RBoxNRSuppression} and follows from taking the limit $k\to\infty$ of the last inequality in equation~\eqref{eq:SuppressionRangesExampleIICaseII}. 

In contrast to the invariants $R\Box^k R$ which only put a lower bound on suppressed values for $\beta$, the invariants $R^n$ also put a second, upper, bound on suppressed spacetimes labelled by $\beta$. For $n>1$ the upper bound lies below the lower bound, cf.~equation~\eqref{eq:SuppressionRangesExampleIICaseII}, and thus the full range $\{\beta \in \mathbb{R}\setminus{\{0\}}\}$ can be suppressed, e.g., by actions of quadratic or higher order in the curvature. 

The impossibility of $R \Box^k R$ terms to render the spacetime integral divergent and thereby achieve dynamical singularity suppression of singular spacetimes can be explained by the action~\eqref{eq:BoxActionExampleIICaseII} of the box operator. If $\beta <-2$, the box operator increases the degree of a homogenous function and thus its effect when applied consecutively to the Ricci scalar given in~\eqref{eq:RicciScalarExampleIICaseII}, is to regularize the divergence. Thus, interestingly, in this example we observe a complementarity between the regularity of polynomial Riemann invariants without covariant derivatives of the metric, such as powers of the Ricci scalar $R^n$, versus curvature-derivative invariants involving explicit covariant derivatives of the metric, such as $R\Box^k R$. Certain spacetimes with a higher-degree curvature singularity as measured by $R^n$ can feature more regular curvature-derivative invariants $R\Box^k R$ at a given order $n=k+2$ of metric differentiations (and similarly, e.g., for $\Box^l R$ with $n=l+1$). 

\subsubsection{Case III\,: $\alpha\neq 0,\,\beta\neq0$}\label{SecSubSub:Example2Case3}

If the two free metric functions are given by $A(r) = r^\alpha$ and $B(r)=r^\beta$ with $\alpha,\beta \neq 0$, the $r$-monomials contained in the invariants $R^n$ and $R\Box^k R$ are still described by~\eqref{eq:ScalarsContentExampleIICaseII}, however, the metric determinant~\eqref{eq:MetricDeterminantExampleII} now depends on both parameters $\alpha$ and $\beta$. By applying~\eqref{eq:IntegralFormula}, the corresponding action contributions can achieve destructive interference respectively of spacetimes in the parameter ranges
\ba\label{eq:SuppressionRangesExampleIICaseIII}
\alpha &\leq  -6 +4n -\beta \,;\,-6 +4n +(2n-1)\beta \,,\,\,\, &n\in \mathbb{N} \nn\\
\alpha &\leq 2 + 4k + (2k-1)\beta \,;\, 2 + 4k + (2k+3)\beta  \,,\,\,\,&k\in \mathbb{N}\,,\nn\\
\ea
where we have again restricted our attention to the minimal and maximal values for $z$ in equation~\eqref{eq:ScalarsContentExampleIICaseII}.

The parameter ranges corresponding to singular and suppressed spacetimes via action contributions $R^n$ and $R\Box^k R$ are shown in Figure~\ref{Fig:RNSuppression} and Figure~\ref{Fig:RBoxNRSuppression} in blue color, whereas red regions mark spacetimes with a curvature singularity which cannot be suppressed by these terms. As illustrated on the example of the subcases I and II, the full suppression can only be guaranteed by polynomial invariants of infinite order in the curvature tensor. The invariants $R\Box^k R$, instead, become less efficient in achieving divergence of the spacetime integral for $\beta<-2$. In particular, in the limit $k\to \infty$ all spacetimes with $\beta < -2$ remain unsuppressed. The infinite sum $\lim_{N\to\infty} \sum_{k=1}^{N}\gamma_k R\Box^k R$ in the action with couplings $\gamma_{1,N}\neq 0$ achieves the suppression of the singular parameter space only up to a wedge bounded from above by $\beta = -6 +\alpha$ and $\beta = -2$, as illustrated by the red region in the fourth panel of Figure~\ref{Fig:RBoxNRSuppression} and can be verified from equation~\eqref{eq:SuppressionRangesExampleIICaseIII}. Indeed, a point $(\alpha,\beta)$ in the interior of this wedge can be parametrized as
\be\label{eq:RedRegionParametrization}
	\begin{pmatrix}
	\alpha\\
	\beta
	\end{pmatrix} = \begin{pmatrix}
	4\\
	-2
	\end{pmatrix}
+  \lambda_1\begin{pmatrix}
	1 \\
	0
\end{pmatrix}
+  \lambda_2\begin{pmatrix}
	-1 \\
	-1
\end{pmatrix}\,, \,\,\,\lambda_{1,2} >0\,.
\ee
Inserting these expressions for $\alpha$ and $\beta$ into the second line of equation~\eqref{eq:SuppressionRangesExampleIICaseIII}, these inequalities are seen to be violated for all $k\in \mathbb{N}$ due to the assumption of positivity of $\lambda_{1,2}$.
Thus singular spacetimes with parameters $(\alpha,\beta)$ in the red region can not be suppressed by combinations of terms $R\Box^k R$ in the action for the same reason as anticipated in the analysis of subcase II in Section~\ref{SecSubSub:Example2Case2}. The d'Alembert operator acting on the Ricci scalar for spacetimes in the red wedge smoothens out the divergence and thereby makes the resulting invariant more regular.
Instead, the singular spacetimes described by this family can be asymptotically suppressed in the path integral only by local higher-curvature terms $R^n$ for $n\to \infty$.

Altogether, the previous results indicate that achieving destructive interference of spacetimes with a curvature singularity in the path integral requires explicitly actions of higher order in the curvature, rather than just in the counting of derivatives of the metric tensor.

\section{Discussion}\label{Sec:Discussion}

Lorentzian path integrals provide a promising framework for quantum gravity. While methods for the evaluation of highly oscillatory integrals are advancing~\cite{Tate:2011ct,Dittrich:2014mxa,Delcamp:2016dqo,Feldbrugge:2017fcc,Feldbrugge:2017kzv,Feldbrugge:2019fjs,Jia:2021xeh,Asante:2021phx,Feldbrugge:2022idb,Loges:2022nuw,Dittrich:2023rcr} and there is accumulating evidence that Lorentzian path integrals evade some of the long-standing debates related to the conformal factor problem in Euclidean quantum gravity~\cite{Giddings:1989ny,Marolf:2022ybi,Borissova:2023izx,Dasgupta:2001ue},
in this work we have pointed out a further property of Lorentzian path integrals as opposed to their Euclidean counterparts: the prospect for suppression of off-shell spacetime geometries with an unphysical curvature singularity. 
The suppression mechanism grounds on the observation that whenever the magnitude of the action is large for a set of neighboring singular off-shell geometries in configuration space, the rapidly oscillating phase factor is expected to lead to destructive inteference between these~\cite{Lehners:2019ibe,Borissova:2020knn, Lehners:2023fud}. Demanding such a property of the Lorentzian path integral translates into a selection principle constraining the microscopic action and thereby the dynamics of gravity.

We have investigated the effects of local corrections $\mathcal{R}^n$ and $\mathcal{R}\Box^k \mathcal{R}$, with $n,k\in \mathbb{N}$, in the action for simple families of static spherically symmetric spacetimes. Our results suggest that including terms $\mathcal{R}^n$ for $n\to \infty$ provides a robust way of achieving a divergent action integral for singular spacetimes on approach to the singularity, and therefore a robust way of suppressing singular spacetimes in the Lorentzian gravitational path integral. By contrast, an analysis of the static spherically symmetric metrics where the two free functions are described by a power law in the radial coordinate, shows that $\mathcal{R}\Box^k \mathcal{R}$ invariants can not reach the full parameter space of singular spacetimes. The latter can be explained by the non-trivial action of the covariant d'Alembert operator in curved spacetimes. Thereby, including higher derivatives in the form of powers of the box operator acting on a curvature may increase the degree of regularity of a scalar invariant and thus smooth out a possible curvature divergence. Altogether, we observe a complementarity between the singularity of curvature invariants and regularity of curvature-derivative invariants in certain singular spacetimes. At the same time, the analysis of the first family of static spherically symmetric spacetimes illustrates that even spacetimes for which all polynomial Riemann invariants $\mathcal{R}^n$ are non-singular, can exhibit a divergence in curvature-derivative invariants $\mathcal{R}\Box^k \mathcal{R}$ beyond a certain $k\in\mathbb{N}$. Related observations were reported already in~\cite{Burzilla:2020utr,Giacchini:2021pmr,Giacchini:2023waa}, where it was pointed out that various regular black-hole spacetimes with smooth de Sitter cores have divergent curvature-derivative invariants. Similarly, in our first example, the effect of including curvature invariants with an increasing number of covariant derivatives is generically to push the bound of suppressed spacetimes to increasingly regular ones. In turn, one might object that in a Lorentzian path integral in which the gravitational action compiles with the requirement of dynamically suppressing all singular spacetimes, necessarily, certain regular spacetimes are suppressed too~\cite{Giacchini:2021pmr}. Therein lies no fundamental obstruction as long as these regular spacetimes are off-shell, given that, ultimately, candidate geometries for quantum spacetime should arise as singularity-free solutions to the quantum dynamics. \footnote{The compatibility of some of the familiar regular black-hole spacetimes admitting an asymptotic expansion at infinity, with a variational action principle, seems to be challenged~\cite{Knorr:2022kqp}.} 

At this point, some further comments are in order. The spacetimes~\eqref{eq:ActionSpherical SymmetryPowerLaw} and the examples investigated in Section~\ref{Sec:ActionPhaseFactor} represent just a subset of all geometries integrated over in the path integral (viewed as a general object object summing over elements out of a predefined configurations space, as emphasized in the Introduction~\ref{Sec:Introduction}). Nevertheless, they are sufficient to arrive at the above conclusions. To see this, it is important to note that the argument put forward here makes use of the logical equivalence of the statements $A\Rightarrow B$ and $\neg B \Rightarrow \neg A$. Accordingly, if it is found that a certain set of singular spacetimes integrated over in the path integral can not interfere away via any finite number of curvatures in the action, or via specific curvature invariants involving covariant d’Alembertians, it follows that an action built solely from these types of invariants can not satisfy the requirement of suppressing all off-shell spacetimes in the path integral. Consequently, such an action must be discarded according to the divergent-action requirement for singularity suppression.

Let us moreover emphasize that the conclusions derived here in the context of spherically symmetric spacetimes~\eqref{eq:MetricSphericalSymmetry} with curvature singularity at $r=0$, apply straightforwardly to cosmological singularities. The difference would be that in the latter case curvature invariants diverge as a function of a given time coordinate, such that the relevant integral in the action would be the integral over this time coordinate.

A further comment concerns the explicit evaluation of a path integral for gravity, in which there are inifintely many configurations with a finite action. For example, the regular spacetimes in the green region of Figure~\ref{Fig:RNSuppression} whose parameter takes values $\alpha>2$, exhibit a finite action as $r\to 0$, provided the latter is built only from curvature invariants, but does not contain curvature-derivative invariants. It should be emphasized that such a situation poses a generic problem, e.g., for path integrals in which the action is a topological invariant, but still geometries with the same topology are integrated over. We did not ambition to solve a problem of this sort here. Instead, the main focus of this work is the possibility of suppressing an increasing range of singular spacetimes in a Lorentzian path integral through increasing orders in curvatures of the action. In this sense this work should be viewed as a step towards combining the properties of Lorentzian path integrals with the active field of study of infinite-curvature and infinite-derivative classical actions, and to point out the possibility that curvature singularities at the classical level may be subject to an intrinsic suppression mechanism at the quantum level.
 Clearly, the viewpoint on quantum gravity via Lorentzian path integrals adopted here, differs from the one taken in minisuperspace analyses~\cite{Hartle:1983ai,Feldbrugge:2017fcc,Feldbrugge:2017kzv,DiazDorronsoro:2017hti,Feldbrugge:2017mbc,DiazDorronsoro:2018wro,Feldbrugge:2018gin} (see, e.g., \cite{Lehners:2023yrj} for a review on the no-boundary wavefunction), which typically consider the dynamics dictated by the Einstein-Hilbert action, and make symmetry assumptions in order to reduce the gravitational path integral to an integral over a finite number of degrees of freedom. In such a framework, an explicit evaluation of the corresponding path integral is possible, at least within saddle-point approximations and taking into account Picard-Lefschetz theory~\cite{Feldbrugge:2017fcc,Lehners:2023yrj} or other criteria such as the coupling to matter~\cite{Witten:2021nzp}, in order to determine which saddle points are relevant. In such contexts, one may extract concrete physical predictions from the corresponding minisuperspace models. Whether or not such minisuperspaces provide a viable approach to study the full gravitational path integral, is an open question. Our considerations here on a Lorentzian path integral for gravity are much more general and independent of the choice of dynamics or approximate methods for evaluation.

The previous observations furthermore raise the fundamental question about the fate of on-shell configurations in the path integral with a divergent action due to the specified boundary conditions. 
An example for gravity is the Schwarzschild spacetime as solution to a quadratic-gravity action $S_\epsilon$ including in particular the Kretschmann scalar $R_{\mu\nu\rho\sigma}R^{\mu\nu\rho\sigma}$ or the squared Weyl tensor $C_{\mu\nu\rho\sigma}C^{\mu\nu\rho\sigma}$, whose spacetime integral diverges as $\epsilon^{-3}$ in the limit $\epsilon\to 0$ where $\epsilon$ represents a regularizing small-distance cutoff.
While partition functions for a fixed volume of space have been recently investigated based on the action of general relativity~\cite{Jacobson:2022jir} and higher-derivative gravity~\cite{Tavlayan:2023sqm}, sheding light on the above question can be done by studying path integrals as function(al)s of variable or dynamical spatial boundary conditions, e.g., in our toy-model example $ Z(\epsilon) \equiv Z(S_\epsilon)$ with varying parameter $\epsilon$. Alternatively, in the context of an analytical-mechanics system, one would have to investigate explicitly solutions with a divergent action resulting, e.g., from a divergent potential at the boundary, and determine their contribution to the Lorentzian path integral using for instance Picard-Lefschetz theory~\cite{Feldbrugge:2017fcc,Feldbrugge:2017kzv}.

In the search for a microscopic action for gravity constrained by the requirement of suppressing off-shell spacetimes with a curvature singularity in the Lorentzian gravitational path integral, our analysis allows us to address in particular actions of the type
\ba\label{eq:ActionFormFactors}
S = S_{\text{EH}} &+& \int \dd[4]{x}\sqrt{-g}\Big(R \mathcal{F}_1(\Box) R + R_{\mu\nu} \mathcal{F}_2(\Box) R^{\mu\nu} \nn\\
&+& R_{\mu\nu\rho\sigma}\mathcal{F}_3(\Box)  R^{\mu\nu\rho\sigma}\Big)\,,
\ea
where the form-factor functions $\mathcal{F}_i$ are semi-local functions in the covariant d'Alembertian admitting a Taylor expansion around $\Box = 0$. These include renormalizable quadratic gravity for $\mathcal{F}_i(\Box) =\gamma_i \mathbb{I}$~\cite{Stelle:1976gc}, superrenormalizable higher-derivative gravity for $\mathcal{F}_i(\Box) = \sum_{n=0}^N \gamma_{i,n} \Box^n$ with $N\in \mathbb{N}$~\cite{Asorey:1996hz} and ghostfree Lee-Wick models for special choices of the coefficients~$\gamma_{i,n}$~\cite{Lee:1969fy,Lee:1970iw,Modesto:2014lga,Modesto:2015ozb,Modesto:2016ofr,Anselmi:2017lia}. Actions of the form~\eqref{eq:ActionFormFactors} are moreover of relevance for the form-factor program in asymptotically safe quantum gravity~\cite{Knorr:2019atm,Knorr:2021iwv,Knorr:2022dsx}. As infinite-derivative expansions they appear in non-local ghostfree gravity~\cite{Modesto:2017sdr,BasiBeneito:2022wux,Biswas:2011ar,Krasnikov:1987yj,Modesto:2011kw,Tomboulis:1997gg}, asymptotically non-local gravity~\cite{Boos:2022biz} and as effective actions in loop quantum gravity~\cite{Borissova:2022clg,Borissova:2023gmj}. The analysis of this paper suggests that actions of the form~\eqref{eq:ActionFormFactors} are incompatible with a dynamical singularity-suppression principle, unless an equivalence to an action of infinite order in the curvature can be established. Clearly, such an equivalence requires a careful treatment of boundary terms and at least the form factors $\mathcal{F}_i$ to contain infinitely many derivatives of the metric. \footnote{While there exist generalizations of the Gauss-Bonnet identity indicating that a certain combination of invariants $\mathcal{R}\Box^k \mathcal{R}$ can be rewritten in terms of invariants of higher order in the curvature and additional contributions, cf.~\cite{Asorey:1996hz,Deser:1986xr}, to the best of our knowledge a general relation between infinite-derivative versus infinite-curvature gravitational actions has not been established.}

 
A compelling question is therefore, whether infinite-derivative and non-local theories, tentatively favored by the principle of singularity suppression, might indeed feature non-singular black-hole solutions. There exist several indications that this is the case. In particular, regular black-hole spacetimes have been found in superrenormalizable infinite-derivative quantum gravity~\cite{Modesto:2011kw,Giacchini:2018zup}, non-local gravity~\cite{Biswas:2011ar,Buoninfante:2020ctr,Giacchini:2018wlf,Giacchini:2023waa,dePaulaNetto:2023vtg}, asymptotically safe quantum gravity~\cite{Bonanno:2000ep,Falls:2010he,Torres:2014gta,Torres:2017ygl,Adeifeoba:2018ydh,Pawlowski:2018swz,Platania:2019kyx,Platania:2023srt,Bonanno:2023rzk} and recent studies explicitly taking into account infinite towers of higher-curvature corrections~\cite{Bueno:2024dgm}. 
These findings contrast black-hole solutions in perturbatively renormalizable models, such as quadratic gravity where spacetime singularities persist~\cite{Stelle:1977ry,Lu:2015psa}. Altogether, imposing a selection-principle of suppressing off-shell singular spacetimes in Lorentzian amplitudes singles out non-perturbative models for quantum gravity. The derivation of full solutions to these non-perturbative models and study of their physical viability as effective spacetime models would provide an important increment to the quantum mechanism of off-shell singularity-suppression in the path integral, based on the infinite-action requirement, and remains a key open challenge.

\acknowledgments{

We thank Bianca Dittrich, Jos\'e Padua-Arg\"uelles and Alessia Platania for inspiring discussions, Luca Buoninfante for written communication, Bianca Dittrich and Alessia Platania for comments on the draft, and Astrid Eichhorn for valuable suggestions on an earlier draft. The author is supported by an NSERC grant awarded to Bianca Dittrich and a research grant by the Blaumann Foundation. Research at Perimeter Institute is supported in part by the Government of Canada through the Department of Innovation, Science and Economic Development Canada and by the Province of Ontario through the Ministry of Colleges and Universities. 

\appendix

\section{Polynomial basis for Riemann invariants}\label{Sec:ZMInvariants}

A complete polynomial basis for the set of polynomial Riemann invariants $\{\mathcal{R}^n\,|\,n\in \mathbb{N}\}$ of an arbitrary four-dimensional spacetime is provided by the Zakhary-McIntosh (ZM) invariants~\cite{Carminati:1991zm,Mcintosh1997:zm,Carminati:2002zm}. These consist of four Weyl invariants $\mathcal{K}^{\text{ZM}}_{1-4}$, four Ricci invariants $\mathcal{K}^{\text{ZM}}_{5-8}$, and nine Ricci-Weyl invariants $\mathcal{K}^{\text{ZM}}_{9-17}$,
\ba
\mathcal{K}^{\text{ZM}}_1 &=& C_{\mu\nu\rho\sigma}C^{\mu\nu\rho\sigma}\,,\nn\\
\mathcal{K}^{\text{ZM}}_2  &=& C_{\mu\nu\rho\sigma}\tilde{C}^{\mu\nu\rho\sigma}\,,\nn\\
\mathcal{K}^{\text{ZM}}_3  &=& C\indices{_\mu_\nu ^\rho ^\sigma} C\indices{_\rho_\sigma^\alpha^\beta}C\indices{_\alpha_\beta^\mu^\nu} \,,\nn\\
\mathcal{K}^{\text{ZM}}_4  &=& \tilde{C}\indices{_\mu_\nu ^\rho ^\sigma} C\indices{_\rho_\sigma^\alpha^\beta}C\indices{_\alpha_\beta^\mu^\nu}\,,\nn\\
\mathcal{K}^{\text{ZM}}_5  &=& R\,,\nn\\
\mathcal{K}^{\text{ZM}}_6  &=& R\indices{_\mu^\nu}R\indices{_\nu^\mu} \,,\nn\\
\mathcal{K}^{\text{ZM}}_7  &=& R\indices{_\mu^\nu}R\indices{_\nu^\rho} R\indices{_\rho^\mu}\,,\nn\\
\mathcal{K}^{\text{ZM}}_8  &=& R\indices{_\mu^\nu}R\indices{_\nu^\rho} R\indices{_\rho^\sigma} R\indices{_\sigma^\mu}\,,\nn\\
\mathcal{K}^{\text{ZM}}_9  &=& R\indices{^\mu^\nu}R\indices{^\rho^\sigma} C_{\mu\nu\rho\sigma}\,,\nn\\
\mathcal{K}^{\text{ZM}}_{10}  &=& R\indices{^\mu^\nu}R\indices{^\rho^\sigma} \tilde{C}_{\mu\nu\rho\sigma}\,,\nn\\
\mathcal{K}^{\text{ZM}}_{11}  &=& R\indices{_\mu ^\alpha} R\indices{_\alpha^\rho}  R\indices{_\nu ^\beta} R\indices{_\beta ^\sigma} C\indices{^\mu^\nu_\rho_\sigma}\,,\nn\\
\mathcal{K}^{\text{ZM}}_{12}  &=& R\indices{_\mu ^\alpha} R\indices{_\alpha^\rho}  R\indices{_\nu ^\beta} R\indices{_\beta ^\sigma} \tilde{C}\indices{^\mu^\nu_\rho_\sigma}\,,\nn\\
\mathcal{K}^{\text{ZM}}_{13}  &=& R\indices{^\mu^\nu} R\indices{_\rho_\sigma}C\indices{_\alpha _\mu_\nu _\beta}\tilde{C}\indices{^\alpha ^\rho^\sigma ^\beta}\,,\nn\\
\mathcal{K}^{\text{ZM}}_{14}  &=& R\indices{^\mu^\nu} R\indices{_\rho_\sigma}\qty(C\indices{_\alpha _\mu_\nu _\beta} C\indices{^\alpha^\rho^\sigma^\beta} + \tilde{C}\indices{_\alpha _\mu_\nu _\beta} \tilde{C}\indices{^\alpha^\rho^\sigma^\beta} )\,,\nn \\
\mathcal{K}^{\text{ZM}}_{15}  &=& R\indices{^\mu^\nu}R\indices{_\rho_\sigma} \qty(C\indices{_\alpha _\mu _\nu _\beta} C\indices{^\alpha ^\rho^\sigma^\beta}-\tilde{C}\indices{_\alpha _\mu _\nu _\beta} \tilde{C}\indices{^\alpha ^\rho^\sigma^\beta})\,,\nn\\
\mathcal{K}^{\text{ZM}}_{16}  &=& R\indices{^\mu^\nu}R\indices{^\rho^\sigma}C\indices{^\alpha ^\beta^\gamma^\delta}\qty(C\indices{_\alpha _\mu_\nu _\delta}C\indices{_\beta_\rho_\sigma_\gamma} + \tilde{C}\indices{_\alpha _\mu_\nu _\delta} \tilde{C}\indices{_\beta_\rho_\sigma_\gamma} )\,,\nn\\
\mathcal{K}^{\text{ZM}}_{17}  &=&R\indices{^\mu^\nu}R\indices{^\rho^\sigma}\tilde{C}\indices{^\alpha ^\beta^\gamma^\delta}\qty(C\indices{_\alpha _\mu_\nu _\delta}C\indices{_\beta_\rho_\sigma_\gamma} - \tilde{C}\indices{_\alpha _\mu_\nu _\delta} \tilde{C}\indices{_\beta_\rho_\sigma_\gamma} )\nn\,,
\ea
(see, e.g., \cite{Overduin:2020aiq,Held:2021vwd}). Here $\tilde{C}$ denotes the dual Weyl tensor $\tilde{C}_{\mu\nu\rho\sigma} = \frac{1}{2}\epsilon_{\mu\nu\alpha\beta}C\indices{^\alpha^\beta_\rho_\sigma}$.

\subsection*{ZM invariants for example I\,: $A(r)=B(r)=1-r^\alpha$}\label{SecSub:ZMInvariantsExampleI}

Table~\ref{Tab:ZMInvariantsExample1} shows the ZM invariants for static spherically symmetric spacetimes with line element~\eqref{eq:MetricSphericalSymmetry} and lapse function~\eqref{eq:ExampleIFunctionA}, computed using the {\scshape Wolfram Mathematica}~\cite{Mathematica:2023wm} tensor-algebra package {\scshape xAct}~\cite{xAct:2023wm}. For generality we have included a possible non-zero real coefficient $a$ in front of $r^\alpha$ and written the lapse function as $A(r)=1-ar^\alpha$. All non-vanishing ZM invariants in this case can be expressed as
\be
\mathcal{K}_i^{\text{ZM}} = {a}^n  p_i^{(n)}(\alpha)  r^{n(\alpha-2)}\,,
\ee
with a polynomial $p_i^{(n)}(\alpha)$ of order $n$ in $\alpha$, whereby $n \in \mathbb{N}$ denotes the number of contracted curvatures appearing in the definition of the ZM invariant. For each spacetime from the family~\eqref{eq:ExampleIFunctionA} characterized by a fixed value of $\alpha$, a single non-vanishing ZM invariant can be selected to express all other ZM invariants as a polynomial thereof. Thus, for a given spacetime labeled by $\alpha$, only one ZM invariant is sufficient to generate all polynomial Riemann invariants of the form $\mathcal{R}^n$ for this spacetime. In particular, any non-vanishing Riemann invariant consisting of $n\in \mathbb{N}$ contracted Riemann curvature tensors takes the form
\be
\mathcal{R}^n \propto r^{n(\alpha-2)} \,.
\ee

One may ask whether there is a curvature invariant which can be used to generate all polynomial Riemann invariants for the entire family of spacetimes~\eqref{eq:ExampleIFunctionA}, i.e., for any value of $\alpha$. This can not be any of the ZM invariants $\mathcal{K}_i^{\text{ZM}}$ in Table~\ref{Tab:ZMInvariantsExample1}, as all the polynomials  $p_i^{(n)}(\alpha)$ exhibit one or more real zeros and therefore there is no ZM invariant which is non-vanishing for all $\alpha$.
However, the Kretschmann scalar
\be
R_{\mu\nu\rho\sigma}R^{\mu\nu\rho\sigma} =a^2  \qty(\alpha^4 - 2\alpha^3 +5\alpha^2 + 4)r^{2(\alpha-2)}
\ee
contains a polynomial of order four in $\alpha$ with no real zeros, i.e., it is non-vanishing for all spacetimes from the one-parameter family~\eqref{eq:ExampleIFunctionA} and can be used to generate all ZM invariants and thereby all polynomial Riemann invariants of the family of spacetimes~\eqref{eq:ExampleIFunctionA}. In this sense the Kretschmann scalar and powers thereof are particularly suited for an action to achieve dynamical suppression of singular spacetimes in the path integral, as there do not appear isolated values of $\alpha$ in the range $\{\alpha < 2\}$, corresponding to singular spacetimes, for which the associated action terms would evaluate to zero. 
It follows furthermore that all polynomial Riemann invariants with $n\in \mathbb{N}$ contracted Riemann curvature tensors can be expressed as
\be
\mathcal{R}_I^{n} = \frac{p_I^{(n)}(\alpha)}{\qty(\alpha^4 - 2\alpha^3 +5\alpha^2 + 4)^{n/2}}\big(R_{\mu\nu\rho\sigma}R^{\mu\nu\rho\sigma}\big)^{n/2}
\ee
with polynomials $p_I^{(n)}(\alpha)$ of order $n$ in $\alpha$. It is clear that in any $\mathcal{R}^n$ invariant the coefficient $a$ appears only as global factor $a^n$ counting the order in curvature, and thus contributes only by proportionality factor to the spacetime integral over such an invariant. We have therefore for simplicity fixed $a=1$ in the main text.

\begin{table}[htb]
	\centering
	\begin{tabular}[t]{c|c}
		\toprule
		{\bf ZM invariant } & $A(r)=B(r)^{-1}=1-ar^\alpha$ \\
		\midrule
		$\mathcal{K}_{1}^{\text{ZM}}$&$\frac{1}{3}a^2 \qty( \alpha^2 -3\alpha+2)^2 r^{2(\alpha-2)}$ \\
		$\mathcal{K}_{2}^{\text{ZM}}$&0\\
		$\mathcal{K}_{3}^{\text{ZM}}$&$\frac{1}{18}a^3\qty(\alpha^2 -3\alpha+2)^3 r^{3(\alpha-2)}$\\
        $\mathcal{K}_{4}^{\text{ZM}}$&0\\
        $\mathcal{K}_{5}^{\text{ZM}}$&$a \qty(\alpha^2+3\alpha+2 ) r^{\alpha-2} $\\
        $\mathcal{K}_{6}^{\text{ZM}}$&$\frac{1}{2}a^2 \qty(\alpha^2+4) \qty(\alpha+1)^2 r^{2(\alpha -2)}$\\
        $\mathcal{K}_{7}^{\text{ZM}}$&$\frac{1}{4}a^3 \qty(\alpha^3 + 8)\qty(\alpha+1)^3 r^{3(\alpha -2)}$\\
        $\mathcal{K}_{8}^{\text{ZM}}$&$\frac{1}{8}a^4 \qty(\alpha^4 + 16)\qty(\alpha+1)^4 r^{4(\alpha -2)}$\\
        $\mathcal{K}_{9}^{\text{ZM}}$&0\\
        $\mathcal{K}_{10}^{\text{ZM}}$&0\\
        $\mathcal{K}_{11}^{\text{ZM}}$&$\frac{1}{48}a^5 \qty(\alpha+1)^4 \qty(\alpha-2)^3 \qty(\alpha+2)^2\qty(\alpha-1)r^{5(\alpha-2)}$\\
        $\mathcal{K}_{12}^{\text{ZM}}$&0\\
        $\mathcal{K}_{13}^{\text{ZM}}$&0\\
        $\mathcal{K}_{14}^{\text{ZM}}$&$\frac{1}{36}a^4\qty(\alpha^2-1)^2 \qty(\alpha-2)^4 r^{4(\alpha-2)}$\\
        $\mathcal{K}_{15}^{\text{ZM}}$&$\frac{1}{36}a^4\qty(\alpha^2-1)^2 \qty(\alpha-2)^4 r^{4(\alpha-2)}$\\
        $\mathcal{K}_{16}^{\text{ZM}}$&$-\frac{1}{108}a^5 \qty(\alpha-2)^5 \qty(\alpha-1)^3 \qty(\alpha+1)^2 r^{5(\alpha -2)}$\\
        $\mathcal{K}_{17}^{\text{ZM}}$&$-\frac{1}{108}a^5 \qty(\alpha-2)^5 \qty(\alpha-1)^3 \qty(\alpha+1)^2 r^{5(\alpha -2)}$\\
		\bottomrule
	\end{tabular}
	\caption{ZM invariants for static spherically symmetric spacetimes of the form~\eqref{eq:MetricSphericalSymmetry} with $A(r)=B(r)^{-1} = 1-ar^\alpha $.}\label{Tab:ZMInvariantsExample1}
\end{table}%

\newpage

\bibliography{References}

\end{document}